\documentclass[conference,10pt]{IEEEtran}
\usepackage{epsfig,rotating,setspace,latexsym,amsmath,epsf,amssymb,amsfonts,bm,theorem,subfigure,epstopdf}
\usepackage{cite,authblk}
\usepackage{bbm}
\usepackage{algorithmic,algorithm}
\newtheorem{theorem}{Theorem}
\newtheorem{lemma}{Lemma}
\newenvironment{Proof}[1]{\medskip\par\noindent{\bf Proof:\,}\,#1}{{\mbox{\,$\blacksquare$}\par}}
\IEEEoverridecommandlockouts

\allowdisplaybreaks
\usepackage{color}
\usepackage{graphicx}

\begin{document}
	
	\title{Age of Information Scaling in Large Networks with Hierarchical Cooperation\thanks{This work was supported by NSF Grants CNS 15-26608, CCF 17-13977 and ECCS 18-07348.}}
	
	\author[1]{Baturalp Buyukates}
	\author[2]{Alkan Soysal}
	\author[1]{Sennur Ulukus}
	\affil[1]{\normalsize Department of Electrical and Computer Engineering, University of Maryland, MD}
	\affil[2]{\normalsize Department of Electrical and Electronics Engineering, Bahcesehir University, Istanbul, Turkey}
	
	\maketitle
	
	\begin{abstract}
		Given $n$ randomly located source-destination (S-D) pairs on a fixed area network that want to communicate with each other, we study the age of information with a particular focus on its scaling as the network size $n$ grows. We propose a three-phase transmission scheme that utilizes \textit{hierarchical cooperation} between users along with \textit{mega update packets} and show that an average age scaling of $O(n^{\alpha(h)}\log n)$ per-user is achievable where $h$ denotes the number of hierarchy levels and $\alpha(h) = \frac{1}{3\cdot2^h+1}$ which tends to $0$ as $h$ increases such that asymptotically average age scaling of the proposed scheme is $O(\log n)$. To the best of our knowledge, this is the best average age scaling result in a status update system with multiple S-D pairs.
	\end{abstract}
	
	\section{Introduction}
	
	Many applications including sensor networking, unmanned aerial vehicle (UAV) systems, and news reports require timely delivery of status update packets that are sent from the source nodes to the interested recipient nodes. In order to measure the freshness of the received information, age of information (AoI) metric has been proposed. Age tracks the time elapsed since the most recent update packet at the destination node was generated at the source node. In other words, at time $t$, age $\Delta(t)$ of a packet which has a timestamp $u(t)$ is $\Delta(t) = t - u(t)$. Age of information has been extensively studied in the literature in a queueing-theoretic setting in references \cite{Kaul12b, Costa14, Huang15, Tripathi19, Bedewy16, Najm17, Soysal18,  Sun16b, soft_upt_allerton, Zou19} and in an energy harvesting setting in references \cite{Yates15, Arafa17b, Arafa17a, Bacinoglu15, Wu18, Arafa18a, Arafa18b, Arafa18d, Baknina18a, Baknina18b}.
	
	With increasing connectivity in communication networks and growing number of information sources (both people and sensors), the issue of scalability of age of information has emerged. In early 00's, followed by the pioneering work of Gupta and Kumar \cite{Gupta00} a similar issue had come up for scaling laws of \emph{throughput} in large networks. References \cite{Gamal06a, Grossglauser02, Neely05, Sharma06} studied throughput scaling in dense and extended networks considering static and mobile nodes. This line of research has culminated in the seminal papers of Ozgur et al. \cite{Ayfer07, Ayfer10} which achieved $O(1)$ throughput per-user by utilizing hierarchical cooperation between nodes. In this paper, we study scaling of \emph{age of information} in large wireless networks.
	
	Recently, the scaling of age of information has been studied in the broadcast setting \cite{Ioannidis09, Zhong17a, Buyukates18, Buyukates18b, Buyukates19}. These works study a single source node which sends status updates to multiple receiver nodes. In \cite{Ioannidis09}, opportunistic contacts between users are utilized to obtain an average  age of $O(\log n)$ at the users. In \cite{Zhong17a, Buyukates18, Buyukates18b, Buyukates19}, single and multihop multicast networks are considered and $O(1)$ average age is obtained at the end nodes by using special transmission schemes such as the earliest $k$ transmission scheme in which the source node waits for delivery to the earliest $k$ out of the total $n$ receiver nodes. Reference \cite{Jiang18a} on the other hand studies age scaling in the multiaccess setting with massive number of source nodes.
	
	In this work, we study a fixed area network of $n$ randomly located source-destination (S-D) pairs that want to send time-sensitive update packets to each other. Each node is both a source and a destination. We aim to find a transmission scheme which allows all $n$ S-D pairs to successfully communicate and achieves the smallest average age scaling per-user.
	
	Reference that is most closely related to our work is \cite{Buyukates18c} in which there are $n$ S-D pairs and an average age scaling of $O(n^\frac{1}{4} \log n)$ per-user is achieved. This work divides the network into cells and utilizes simultaneous transmissions among these cells, provided that the destructive interference caused by other simultaneously active cells is limited, and successive inter-cell MIMO-like transmissions of mega update packets which contain the updates of all nodes from a cell.
	
	In this paper, considering all these previous results, we propose a three-phase transmission scheme that utilizes hierarchical cooperation between users along with mega update packets to serve all S-D pairs. We again divide the network into cells of $M$ nodes each. First and third phases involve intra-cell transmissions and can be performed in parallel across different cells. Second phase, on the other hand, is performed for each cell successively. We observe that the first and third phases essentially require successful communication between pairs but among $M$ nodes rather than $n$. With this observation and the fact that the system is scale-invariant, we can invoke hierarchy in Phases I and III. In other words, we can further divide cells into smaller subcells and apply the proposed three-phase transmission scheme again in Phases I and III. Although hierarchical cooperation was shown to result in poor delay performance in \cite{Ayfer10}, by utilizing mega update packets better age scaling can be achieved here. In fact, using this scheme, we show that an age scaling of $O\left(n^{\alpha(h)}\log n\right)$ per-user is achievable where $\alpha(h) = \frac{1}{3\cdot2^h+1}$ and $h$ denotes the number of hierarchy levels. We note that the scaling result of \cite{Buyukates18c} is the case when hierarchical cooperation is not utilized, i.e., $h=0$. In the asymptotic case when $h \rightarrow \infty$, the proposed scheme achieves an average age scaling of $O(\log n)$.
	
	\section{System Model and Age Metric} \label{model}
	
	We have $n$ nodes that are uniformly and independently distributed on a square of fixed area $S$. These nodes are randomly paired with each other to form $n$ S-D pairs. Each node is both a source and a destination. Each source wants to transmit time-sensitive status update packets to its destination. To measure the freshness of these status update packets, we use the age of information metric. Age is measured for each destination node such that for node $i$ at time $t$ age is the random process $\Delta_i(t) = t - u_i(t)$ where $u_i(t)$ is the timestamp of the most recent update at that node. The metric we use, time averaged age, is given by, $\Delta_i = \lim_{\tau\to\infty} \frac{1}{\tau} \int_{0}^{\tau} \Delta_i(t) dt$ for node $i$. We use a graphical average age analysis to derive the average age for a single S-D pair assuming ergodicity.
	
	The transmission scheme proposed in \cite{Buyukates18c} involves clustering $n$ nodes into $\frac{n}{M}$ cells each with $M$ users and utilizing mega update packets.  In this scheme, the bottleneck in average age scaling is $M$ since during intra-cell transmissions $M$ transmissions are needed, one for each node in a cell. Noting that each cell is a scaled-down version of the whole system, we propose introducing hierarchy by forming subcells from the cells and applying the three-phase transmission scheme on a cell level. Our proposed transmission scheme, therefore, includes inter-cell, inter-subcell (within cells) and intra-subcell transmissions when $h=1$ hierarchy level is utilized. We first analyze the case with $h=1$ level of hierarchy and then generalize the result to $h$ hierarchy levels.
	
	We model the delay in communications between cells as i.i.d.~exponential with parameter $\lambda_0$; between subcells within cells as i.i.d.~exponential with parameter $\lambda_1$ and lastly within subcells as  i.i.d.~exponential with parameter $\lambda_2$.
	
	Due to i.i.d.~nature of service times in these three types of communications, all destination nodes experience statistically identical age processes and will have the same average age. Thus, we will drop user index $i$ in the average age expression and use $\Delta$ instead of $\Delta_i$ in the following analysis.
	
	Finally, we denote the $k$th order statistic of random variables $X_1, \ldots ,X_n$ as $X_{k:n}$. Here, $X_{k:n}$ is the $k$th smallest random variable, e.g., $X_{1:n}=\min\{X_i\}$ and $X_{n:n}=\max\{X_i\}$. For i.i.d.~exponential random variables $X_i$ with parameter $\lambda$,
	\begin{align}
	E [X_{k:n}] =& \frac{1}{\lambda}(H_n - H_{n-k}) \label{ordfirst}\\
	Var[X_{k:n}] =& \frac{1}{\lambda^2}(G_{n} - G_{n-k})
	\end{align}
	where $H_n = \sum_{j=1}^{n} \frac{1}{j}$ and $G_n = \sum_{j=1}^{n} \frac{1}{j^2}$. Using these,
	\begin{align}
	E [X_{k:n}^2] =&  \frac{1}{\lambda^2}\left((H_n - H_{n-k})^2 + G_{n} - G_{n-k} \right) \label{ordlast}
	\end{align}
	
	\section{Age Analysis of a Single S-D Pair} \label{subsection:age}
	
	The proposed scheme involves sessions such that during each session all $n$ S-D pairs are served. Session duration is denoted by random variable $Y$. Here, we derive the average age of a single S-D pair since each pair experiences statistically identical age as explained in Section~\ref{model}.
	
\begin{figure}[t]
		\centering  \includegraphics[width=0.75\columnwidth]{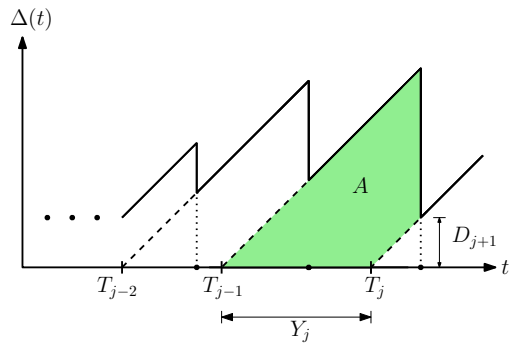}
		\caption{Sample age $\Delta(t)$ evolution for a single S-D pair. Update deliveries  are shown with $\bullet$. Session $j$ starts at time $T_{j-1}$ and lasts until $T_j = Y_j + T_{j-1}$.}
		\label{fig:ageEvol}
		\vspace{-0.5cm}
	\end{figure}

	Session $j$ starts at time $T_{j-1}$ and all source nodes generate their $j$th update packets. This session lasts until time $T_j = T_{j-1}+Y_j$, at which, all $n$ packets are received by their designated recipient nodes. Thus, in the proposed scheme every destination node but one receive their packets before the session ends. Fig.~\ref{fig:ageEvol} shows the evolution of the age at a destination node over time. Upon completion of session $j$ the process repeats itself with session $j+1$.
	
	Using Fig.~\ref{fig:ageEvol} and noting that $Y_j$ and $D_{j+1}$ are independent, as in \cite{Buyukates18c}, the average age for an S-D pair is given by
	\begin{align}
	\Delta =& E[D]+ \frac{E [Y^2]}{2E[Y]}  \label{age_formula}
	\end{align}
	where $D$ denotes the time interval between the generation of an update and its arrival at the destination node. For ease of exposition, we assume that every node updates its age at the end of each session and take $D_{j+1} = Y_{j+1}$ which yields
	\begin{align}
	\Delta =& E [Y] + \frac{E [Y^2] }{2E [Y]} \label{age_propScheme}
	\end{align}
	Note that this assumption can only result in a higher average age as all nodes but one receive their update packets before the session ends, i.e., $D \leq Y$ for all updates and nodes.
	
	\section{Proposed Hierarchical Transmission Scheme} \label{section:scheme}
	
	\begin{figure*}[t]
		\centering  \includegraphics[width=1.60\columnwidth]{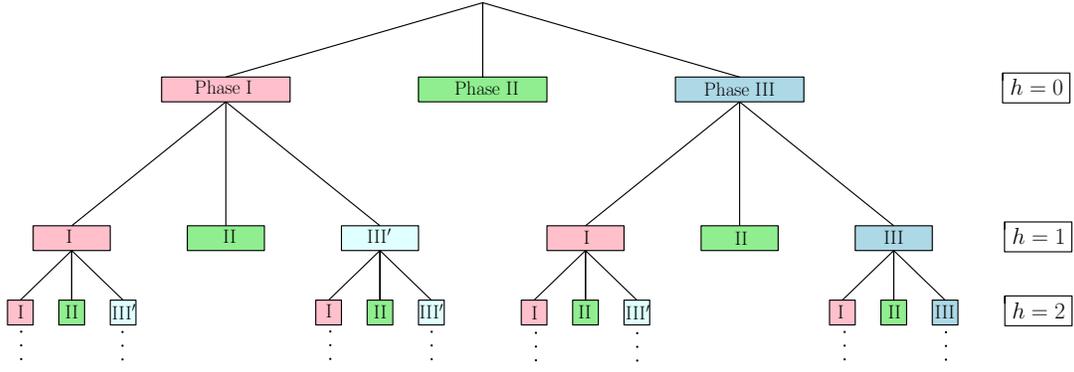} 
		\caption{Proposed three-phase hierarchical transmission scheme.}
		\label{fig:hierarchy}
		\vspace{-0.5cm}
	\end{figure*}
	
	\subsection{Outline of the Scheme} \label{brief}
	A three-phase transmission scheme is proposed in \cite{Buyukates18c} which allows successful communication of $n$ S-D pairs. In that work, the network is divided into $\frac{n}{M}$ cells of equal area such that each cell includes $M$ nodes. In Phase I of the scheme in \cite{Buyukates18c}, nodes in a cell communicate locally to form their mega update packet which includes all $M$ update packets to be sent out from that cell. This phase is simultaneously performed among cells. Then, in Phase II cells operate one at a time to transmit these mega update packets with MIMO-like transmissions to the corresponding cells in which the destination nodes are located. Finally, in Phase III individual packets are extracted from mega update packets and relayed to their respective destination nodes within cells.
	
	Following the analysis to obtain the average age expression by substituting the first and second order moments of the phase durations given in \cite[eqns. (14)-(20)]{Buyukates18c} into the average age expression given in \cite[eqn. (6)]{Buyukates18c} we observe that the resulting per-user average age scaling, when $n$ is large, with $M = n^b$ where $0 < b \leq 1$ and exponential link delays, is characterized by the expected scaling of the phases. As derived in \cite[eqns. (14)-(16)]{Buyukates18c} expected durations of the phases are $O(n^b \log n)$, $O(n^{1-3b} \log n)$ and $O(n^b \log n)$ which in turn result in an average age scaling of $O(n^{\frac{1}{4}} \log n)$ upon selecting $b = 1-3b$. Thus, to obtain a better average age scaling we need to improve the expected length of each phase. This motivates the hierarchical cooperation in the proposed scheme.
	
	In the Phase I of \cite{Buyukates18c}, the communication takes place in between $M = n^b$ nodes rather than $n$ nodes and a simple TDMA operation is performed among these nodes which results in an average scaling of $O(n^b \log n)$. Instead, we introduce the first level of hierarchy by dividing each of these cells into $n^{b-a}$ further subcells with $n^a$ users each where $0<a<b$. Then, we apply the same three-phase scheme with one difference to this cell to accommodate Phase I transmissions of \cite{Buyukates18c}. In particular, to create the mega packet of the cell, first local communication among the nodes is performed within subcells and MIMO-like transmissions are carried out in between subcells within a cell. Then, instead of relaying the received packet to a single node as in Phase~III of \cite{Buyukates18c}, received packets are relayed to every other node in the subcell to create the mega update packet. With this operation, Phase I of \cite{Buyukates18c} is completed in three phases, Phase I, Phase II, Phase III$'$, each of which are scaled down versions of the overall scheme with the corresponding difference in the third phase which is denoted as Phase III$'$ to highlight this difference. The expected length of the first phase with $h=1$ level of hierarchy is then $O(n^a \log n) + O(n^{b-3a} \log n) + O(n^{b-a} \log n)$ (see Section \ref{details} for a detailed derivation) all of which are smaller than $O(n^b \log n)$ achieved in \cite{Buyukates18c}.
	
	Similarly, Phase III of \cite{Buyukates18c} can also be completed in three phases under $h=1$ level of hierarchy. However, this time in the third step we need Phase III rather than Phase III$'$ since we need to relay the received update packet within subcell to its intended recipient node to conclude the delivery.
	
	Fig.~\ref{fig:hierarchy} shows the proposed hierarchy structure in which Phases I and III of level $h$ can be performed by applying the three-phase scheme on a smaller scale at level $h+1$ accordingly. The advantage of the hierarchical transmission is summarized in Table \ref{table:exp_len} with the omission of the common $\log n$ scaling factor in the expressions.
	
	\begin{table}[h]
		\begin{center}
			\begin{tabular}{ | l | l | l |}
				\hline
				& $h=0$\cite{Buyukates18c} & $\hspace{5em}h=1$ \\ \hline
				$\text{Phase I}$ & $O(n^b)$ & $O(n^a)$ + $O(n^{b-3a})$ + $O(n^{b-a})$   \\ \hline
				$\text{Phase II}$  & $O(n^{1-3b})$ & $O(n^{1-3b})$ \\ \hline
				$\text{Phase III}$  & $O(n^b)$ & $O(n^a)$ + $O(n^{b-3a})$ + $O(n^{a})$   \\\hline
			\end{tabular}
		\end{center}
		\caption{Comparison of the expected durations of the phases with $h=0$ \cite{Buyukates18c} and $h=1$ hierarchy level with $0 < a < b \leq 1$. }
		\label{table:exp_len}
		\vspace{-0.75cm}
	\end{table}
	
	\subsection{Detailed Description of the Scheme for $h=1$} \label{details}
	In this section, we describe the proposed hierarchical transmission scheme with $h=1$ level of hierarchy in detail. Later, we generalize the average age scaling result for $h>1$ levels of hierarchy using the fact that the system is scale-invariant. As in \cite{Buyukates18c}, we start with a square network that is divided into $\frac{n}{M}$ cells of equal area with $M$ nodes in each cell with high probability that tends to $1$ as $n$ increases. Selecting $M = n^b$ where $ 0 < b \leq 1$ results in $n^{1-b}$ equal area cells with $n^b$ users each cell. Introducing the first level of hierarchy, we further divide each cell into $n^{b-a}$ equal area subcells to get a total of $n^{1-a}$ subcells with $n^a$ nodes each where  $0 < a < b$.
	
	Transmission delays between the nodes from different cells are denoted by $X^{(0)}_i$, between the nodes from different subcells within the same cell are denoted by $X^{(1)}_i$, and between the nodes belonging to the same subcell are denoted by $X^{(2)}_i$. Note that $X^{(j)}_i$ are independent; and $X^{(j)}_i$ are i.i.d.~exponential with parameter $\lambda_j$ for $j=0,1,2$.
	
	{\bf Phase~I. Creating mega update packets:} In this phase, each cell generates its mega update packet which includes all $M = n^b$ messages to be sent from that cell. As opposed to \cite{Buyukates18c}, we create mega update packets in three successive phases by applying the three-phase transmission scheme to each cell.
	
	First, each node in a subcell distributes its update packet to remaining $n^a-1$ nodes in its subcell which takes $U^{I} = X^{(2)}_{n^a\!-\!1:n^a\!-\!1}$ units of time. Considering $n^a$ successive transmissions for each node of the subcell, this step is completed in a subcell in $V^{I} = \sum_{i=1}^{n^a} U^{I}_i$ units of time. This operation is analogous to the Phase I in \cite{Buyukates18c} but performed among $n^a$ nodes in a subcell rather than among $n^b$ nodes within a cell. Upon completion of this step in a subcell, each node of that subcell has $n^a$ different update packets one from each node. Each node combines all these update packets to form a \emph{preliminary} mega update packet which includes all $n^a$ messages to be sent out from this subcell. This operation is performed in parallel among all subcells in the network (see Section~\ref{note_phaseI} for a detailed description of this operation) and ends when the slowest simultaneously operating subcell finishes creating its \emph{preliminary} mega update packet. Hence, it takes $Y^{I}_{I} = V^{I}_{n^{1\!-\!a}:n^{1\!-\!a}}$ units of time, where $Y^{I}_{I}$ denotes the duration of the first phase at $h=1$.
	
	When all \emph{preliminary} mega update packets are formed, all $n^{b-a}$ subcells of a cell perform MIMO-like transmissions among each other to distribute their \emph{preliminary} mega update packets to remaining subcells within the cell. Since this requires cell-level transmissions in between subcells, this step is performed in parallel among cells and thus, subcells take turns. As in the Phase~II of \cite{Buyukates18c}, all $n^a$ nodes of a subcell start transmitting the \emph{preliminary} mega update packet to remaining $n^{b-a}-1$ subcells. Since every node sends the same \emph{preliminary} mega update packet this does not create interference. This transmission continues until the earliest node in each remaining subcell receives the \emph{preliminary} mega update packet. Thus, for a single subcell it takes $U^{II} = (X^{(1)}_{1:n^{2a}})_{n^{b\!-\!a}\!-\!1:n^{b\!-\!a}\!-\!1}$ units of time. Since subcells take turns, in a cell this step is completed in $V^{II} = \sum_{i=1}^{n^{b-a}} U^{II}_i$ units of time. Finally, on the network-level these MIMO-like transmissions continue until the slowest of the simultaneously operating cells finishes which corresponds to $Y^{II}_{I} = V^{II}_{n^{1\!-\!b}:n^{1\!-\!b}}$.
	
	By the end of the MIMO-like transmissions among subcells, each subcell receives \emph{preliminary} mega update packets of remaining $n^{b-a}-1$ subcells that lie in its cell. In this step, these packets are distributed within the subcell in parallel among the subcells of the network. This is identical to the operation of Phase~III of \cite{Buyukates18c} on subcell-level except that each \emph{preliminary} mega update packet received is transmitted to all nodes of that subcell to successfully form the mega update packet of the corresponding cell. To highlight this difference we denote this step as Phase~III$'$ in Fig.~\ref{fig:hierarchy} at $h=1$ level. Distributing one \emph{preliminary} mega update packet takes $U^{III'} = X^{(2)}_{n^a\!-\!1:n^a\!-\!1}$ units of time. By repeating this for all \emph{preliminary} mega update packets received this step is completed in a subcell in $V^{III'} = \sum_{i=1}^{n^{b\!-\!a}\!-\!1} U^{III'}_i$ units of time. We wait for the slowest subcell and thus on the network-level this step is completed in $Y^{III'}_{I} = V^{III'}_{n^{1\!-\!a}:n^{1\!-\!a}}$ units of time.
	
	With this, each node in a subcell receives remaining $n^{b-a}-1$ \emph{preliminary} mega update packets of $n^a$ message each. Combining these with their own \emph{preliminary} mega update packet, every node in a subcell forms the mega update packet which includes all $n^b$ messages to be sent out from that cell. Thus, the first phase lasts for $Y_I = Y^{I}_I+Y^{II}_I+Y^{III'}_I$ units of time. 
	
	{\bf Phase~II. MIMO-like transmissions:} Identical to Phase~II of \cite{Buyukates18c}, in this phase each cell successively performs MIMO-like transmissions using the mega update packets created in Phase~I. This phase requires network-level transmissions between cells. Thus, only one cell operates at a time. As in \cite{Buyukates18c}, a source node $s$ from cell $j$ needs $\tilde{U} = X^{(0)}_{1:n^{2b}}$ units of time to send its update to the destination cell where the destination node $d$ lies in. Transmissions of cell $j$ continues until all $n^b$ destination cells receive the mega update packet. Hence, for each cell, this phase lasts for $\tilde{V} = \tilde{U}_{n^b:n^b}$. This operation is repeated for each cell and hence the session time of this phase $Y_{II} =\sum_{i=1}^{n^{1-b}} \tilde{V}_i$. At the end of this phase, each cell delivers its mega update packet to one node in each of the corresponding destination cells.
	
	{\bf Phase~III. In-cell relaying to the destination nodes:} By the end of Phase~II, each cell receives a total of $n^b$ mega update packets, one for each node. In \cite{Buyukates18c}, relevant packets which have a destination node in that cell are extracted from these mega update packets and relayed to their respective designated recipient nodes by a simple TDMA operation which scales as $O(n^b \log n)$. However, as in Phase~I we can introduce hierarchy to this phase and apply the three-phase scheme again. Thus, extracted relevant packets are first distributed within subcells of the nodes which received them in Phase~II. Then, these packets are delivered to their corresponding destination subcells in which the destination nodes are located through MIMO-like transmissions and finally, they are relayed to the corresponding recipient nodes within subcells.
	
	Noting that each subcell receives on average $n^a$ mega update packets, with one relevant packet each, distribution of these $n^a$ packets within subcell takes $\hat{V} ^{I} = \sum_{i=1}^{n^a} \hat{U}^{I}_i$ with $\hat{U}^{I} = X^{(2)}_{n^a\!-\!1:n^a\!-\!1}$ and on the network-level is completed in $Y^{I}_{III} = \hat{V}^{I}_{n^{1\!-\!a}:n^{1\!-\!a}}$ units of time. With this operation, the \emph{secondary} mega update packet of that subcell is formed which includes all $n^a$ update packets with destinations in that cell. Then, these \emph{secondary} mega update packets are transmitted to the respective destination subcells in parallel among cells (subcells take turns) through MIMO-like transmissions until all $n^a$ destination cells receive them. In a cell, this is completed in $\hat{V}^{II} = \sum_{i=1}^{n^{b-a}} \hat{U}^{II}_i$ units of time where $\hat{U}^{II} = (X^{(1)}_{1:n^{2a}})_{n^{a}:n^{a}}$ and therefore, on the network-level is completed in $Y^{II}_{III} = \hat{V}^{II}_{n^{1\!-\!b}:n^{1\!-\!b}}$ when all cells finish. Thus, each subcell receives a total of $n^a$ \emph{secondary} mega update packets each of which includes one update destined to a node in that subcell. Finally, these packets are relayed to their actual recipient nodes within subcell. For a subcell it takes $\hat{V}^{III} = \sum_{i=1}^{n^a} \hat{U}^{III}_i$ units of time where $\hat{U}^{III} = X^{(2)}$  and hence on the network-level it is completed in $Y^{III}_{III} = \hat{V}^{III}_{n^{1\!-\!a}:n^{1\!-\!a}}$. Note that since in the last step we relay the packets to their destination node rather than all nodes in the subcell, this step is the subcell-level equivalent of Phase~III of \cite{Buyukates18c}. As a result, the third phase lasts for $Y_{III} = Y^{I}_{III}+Y^{II}_{III}+Y^{III}_{III}$ and finishes when every S-D pair of the network is served.
	
	Total session time of the proposed scheme is, therefore, $Y = Y_I + Y_{II} + Y_{III} $. Before we perform the explicit age calculation using (\ref{age_propScheme}), we make some observations to simplify our analysis.
	\begin{lemma} \label{lemma1}
		$Y_I$ satisfies the following inequality,
		\begin{align}
		Y_I \leq \bar{V}^{I} + \bar{V}^{II} + \bar{V}^{III'} \label{age_ineq}
		\end{align}
		where
		\begin{align}
		\bar{V}^{I} &= \sum_{i=1}^{n^a}\bar{U}^{I}_i,&  &\bar{U}^{I} = X^{(2)}_{n:n} \\
		\bar{V}^{II} &= \sum_{i=1}^{n^{b-a}}\bar{U}^{II}_i,&   &\bar{U}^{II} = (X^{(1)}_{1:n^{2a}})_{n^{1\!-\!a}:n^{1\!-\!a}} \\
		\bar{V}^{III'} &= \sum_{i=1}^{n^{b-a}}\bar{U}^{III'}_i,& &\bar{U}^{III'} = X^{(2)}_{n:n}
		\end{align}
	\end{lemma}
	
	The proof of this lemma follows similarly from that of \cite[Lemma 1]{Buyukates18c}. We show that $Y^{I}_{I} \leq \bar{V}^{I}$, $Y^{II}_I \leq \bar{V}^{II}$ and $Y^{III'}_I \leq \bar{V}^{III'}$ which yields (\ref{age_ineq}).
	
	We worsen our scheme in terms of session time and hereafter take the upper bound in Lemma~\ref{lemma1} as our Phase~I duration for tractability and ease of calculation. Thus, from now on $Y_I = \bar{V}^{I} + \bar{V}^{II} + \bar{V}^{III'} $. Next, we have the following upper bound for the duration of Phase~III.
	
	\begin{lemma} \label{lemma2}
		$Y_{III}$ satisfies the following inequality,
		\begin{align}
		Y_{III} \leq \bar{\bar{V}}^{I} + \bar{\bar{V}}^{II}  + \bar{\bar{V}}^{III} \label{age_ineq2}
		\end{align}
		where
		\begin{align}
		\bar{\bar{V}}^{I} &= \sum_{i=1}^{n^a}\bar{\bar{U}}^{I}_i,&  &\bar{\bar{U}}^{I} = X^{(2)}_{n:n} \\
		\bar{\bar{V}}^{II}  &= \sum_{i=1}^{n^{b-a}}\bar{\bar{U}}^{II}_i,&  &\bar{\bar{U}}^{II} = (X^{(1)}_{1:n^{2a}})_{n^{1\!-\!b\!+\!a}:n^{1\!-\!b\!+\!a}} \\
		\bar{\bar{V}}^{III}  &= \sum_{i=1}^{n^{a}}\bar{\bar{U}}^{III}_i,&  &\bar{\bar{U}}^{III} = X^{(2)}_{n^{1-a}:n^{1-a}}
		\end{align}
	\end{lemma}
	
	We omit the proof of Lemma~\ref{lemma2} since it follows similar to the proof of Lemma~\ref{lemma1}. We worsen Phase~III as well in terms of duration and take $Y_{III} = \bar{\bar{V}}^{I} + \bar{\bar{V}}^{II}  + \bar{\bar{V}}^{III}$ from now on because of similar tractability issues.
	
	As a result of Lemmas \ref{lemma1} and \ref{lemma2}, total session time becomes
	\begin{align}
	Y = \bar{V}^{I} + \bar{V}^{II} + \bar{V}^{III'} + Y_{II} + \bar{\bar{V}}^{I} + \bar{\bar{V}}^{II}  + \bar{\bar{V}}^{III} \label{sessiontime2}
	\end{align}
	
	Taking expectations of order statistics of exponential random variables as in (\ref{ordfirst})-(\ref{ordlast}) and using the fact that for large $n$, we have $H_n \approx \log n$ and $G_{n}$ is monotonically increasing and converges to $\frac{\pi^2}{6}$, first two moments of the subphase and phase durations approximately become
	
	\begin{align}
	E \left[\sum_{i \in \mathcal{I'}} \bar{V}^{(i)}\right] =& \left(\frac{n^a + n^{b-a}}{\lambda_2} \!+\! \frac{(1-a)n^{b-3a}}{\lambda_1}\right) \log n \label{momentfirst_large}\\
	E \left[\sum_{i \in \mathcal{I}} \bar{\bar{V}}^{(i)}\right] =&\left(\frac{(2\!-\!a) n^a}{\lambda_2} \!+\! \frac{(1\!-\!b\!+\!a)n^{b-\!3a}}{\lambda_1}\right)\log n  \label{momentfirst_large2}\\
	E[Y_{II}] =& \frac{bn^{1-3b}}{\lambda_0}\! \log n \\
	E\left[Y_{II}^2\right] =&  \frac{n^{1-5b}}{{\lambda^2_0}}\frac{\pi^2}{6} + \frac{b^2n^{2(1-3b)}}{{\lambda^2_0}}\log^2 n \\
	E\left[\left(\bar{V}^{I}\right)^2\right] =& \frac{n^a}{\lambda_2^2}\frac{\pi^2}{6} + \frac{n^{2a}}{\lambda_2^2} \log^2 n\\
	E\left[\left(\bar{V}^{II}\right)^2\right] =& \frac{n^{b-5a}}{\lambda_1^2}\frac{\pi^2}{6} + \frac{(1-a)^2n^{2(b-3a)}}{\lambda_1^2} \log^2 n\\
	E\left[\left(\bar{V}^{III'}\right)^2\right] =& \frac{n^{b-a}}{\lambda_2^2}\frac{\pi^2}{6} + \frac{n^{2(b-a)}}{\lambda_2^2} \log^2 n\\
	E\left[\left(\bar{\bar{V}}^{I}\right)^2\right] =& \frac{n^a}{\lambda_2^2}\frac{\pi^2}{6} + \frac{n^{2a}}{\lambda_2^2} \log^2 n\\
	E\left[\left(\bar{\bar{V}}^{II}\right)^2\right] =& \frac{n^{b-5a}}{\lambda_1^2}\frac{\pi^2}{6} \!+\! \frac{(1\!-\!b\!+\!a)^2n^{2(b-3a)}}{\lambda_1^2}\! \log^2 n\\
	E\left[\left(\bar{\bar{V}}^{III}\right)^2\right] =& \frac{n^{a}}{\lambda_2^2}\frac{\pi^2}{6} + \frac{(1-a)^2n^{2a}}{\lambda_2^2} \log^2 n\label{momentlast_large}
	\end{align}
	where in (\ref{momentfirst_large}), $ i \in \mathcal{I'} = \{ I, II, III' \}$ and in (\ref{momentfirst_large2}), $ i \in \mathcal{I} = \{ I, II, III \}$ . Now, we are ready to derive an age expression using the above results in (\ref{age_propScheme}).
	
	\begin{theorem} \label{thm1}
		Under the constructed transmission scheme with $h=1$ level of hierarchy, for large $n$, the average age of an S-D pair is given by,
		\begin{align}
		\Delta =& E \left[\sum_{i \in \mathcal{I'}} \bar{V}^{(i)}\right] + E  [Y_{II}] + E  \left[\sum_{i \in \mathcal{I}} \bar{\bar{V}}^{(i)}\right] \nonumber \\ &+ \frac{E \left[\left(\sum_{i \in \mathcal{I'}} \bar{V}^{(i)} + Y_{II} + \sum_{i \in \mathcal{I}} \bar{\bar{V}}^{(i)}\right)^2\right]}{2\left(E  \left[\sum_{i \in \mathcal{I'}} \bar{V}^{(i)}\right] + E  [Y_{II}] + E  \left[\sum_{i \in \mathcal{I}} \bar{\bar{V}}^{(i)}\right]\right)} \label{age_thm}
		\end{align}
	\end{theorem}
	
	The proof of Theorem \ref{thm1} follows upon substituting (\ref{sessiontime2}) back in (\ref{age_propScheme}). Moments follow from (\ref{momentfirst_large})-(\ref{momentlast_large}).
	
	\begin{theorem} \label{therom_ab}
		In order to obtain the minimum scaling for $h=1$ hierarchy level, $b=2a$ needs to be satisfied. This yields an average age scaling of $O(n^{\frac{1}{7}} \log n)$ per-user.
	\end{theorem}
	
	\begin{Proof}
		Using (\ref{momentfirst_large})-(\ref{momentlast_large}) in (\ref{age_thm}), we observe that in the average age expression we have terms with $O(n^a \log n)$, $O(n^{b-a}\log n)$, $O(n^{b-3a}\log n)$, and $O(n^{1-3b}\log n)$. Among first three types, noting that $b-3a < b-a$, dominating terms are $O(n^a)$ and $O(n^{b-a})$. Thus, by choosing $a = b-a$ we can minimize the resulting scaling. With this selection, the first and third terms in (\ref{age_thm}) are $O(n^a\log n)$ whereas the second one is $O(n^{1-6a}\log n)$. Looking at the fourth term we see that the numerator is $O(n^{2a}\log^2 n)$ and the denominator is $O(n^{a}\log n)$ making this term again $O(n^{a}\log n)$. Thus, when we select $a  = \frac{1}{7}$ from $a = 1-6a$ we obtain $O(n^{\frac{1}{7}} \log n)$.
	\end{Proof}
	
	Thus, to get the minimum scaling we need to select $a  = \frac{1}{7}$  and $b  = \frac{2}{7}$. This implies that if the cells have $M$ nodes each, each subcell has $\sqrt{M}$ nodes when $h=1$. Note that in \cite{Buyukates18c} it is shown that $\frac{1}{4} \leq b \leq 1$. Our resulting $b$ not only satisfies this but also gives a better scaling in the end because of the hierarchy we utilized. In Theorem~\ref{thm3} below, we generalize this scaling result to $h$ levels of hierarchy.
	
	\begin{theorem} \label{thm3}
		For large $n$, when the proposed scheme is implemented with $h$ hierarchy levels, the average age scaling of $O\left(n^{\alpha(h)}\log n\right)$ per-user is achievable where $\alpha(h) = \frac{1}{3\cdot2^h+1}$.
	\end{theorem}
	\begin{Proof}
		We observe that when $h=1$ hierarchy level is utilized, the scaling result comes from $a = 1-6a$. Since $b=2a$, another way to express this is $\frac{b}{2^h} = 1-3b$. As $h$ increases with $b=2a$ structure in each level of hierarchy, we see that subcells at level $h$ have $\frac{b}{2^h}$ nodes. However, second phase is still $O(n^{1-3b})$ as each cell at the top of the hierarchy still has $n^b$ nodes. Thus, $\frac{b}{2^h} = 1-3b$ yields $\alpha(h) = \frac{1}{3\cdot2^h +1}$.
	\end{Proof}
	
	Thus, the proposed transmission scheme, which involves hierarchical cooperation and MIMO-like inter-cell transmissions, allows the successful communication of $n$ S-D pairs, and achieves an average age scaling of $O\left(n^{\alpha(h)}\log n\right)$ per-user where $h = 0,1, \ldots$ is the number of hierarchy levels. Note that in the asymptotic case when $h$ tends to $\infty$, the proposed scheme gives an average age scaling of $O(\log n)$ per-user.
	
	\section{Note on Phases~I and~III} \label{note_phaseI}
	
	We model the interference using the protocol model introduced in \cite{Gupta00} such that two nodes can be simultaneously active if they are sufficiently spatially separated. In order for node $j$ to receive an update from node $i$, the following needs to be satisfied for any other node $k$ that is simultaneously active
	\begin{align}
	d(j,k) \geq (1+\gamma)d(j,i) \label{prot_model}
	\end{align}
	where function $d(x,y)$ denotes the distance between nodes $x$ and $y$ and $\gamma$ is the positive guard zone constant.
	
	The proposed hierarchical transmission scheme with $h=1$ level of hierarchy includes within subcell transmissions that are parallelized across subcells and within cell transmissions that are parallelized among cells (subcells take turns) in Phases~I and~III. To implement these parallel subcell-~and cell-level transmissions, we follow a 9-TDMA scheme as in \cite{Ayfer07}. In particular, during cell-level transmissions $\frac{n}{9M}$ of the total $\frac{n}{M}$ cells work simultaneously and during subcell-level transmissions $\frac{n}{9\sqrt{M}}$ of the total $\frac{n}{\sqrt{M}}$ subcells work in parallel. \cite{Buyukates18c} shows that 9-TDMA operation among cells satisfies (\ref{prot_model}) with $\gamma \leq \sqrt{2}-1$. Under the same condition, parallel 9-TDMA operation among subcells is still allowed since from cell-level to subcell-level both distance terms in (\ref{prot_model}) decrease proportionally.  Noting that 9 here is a constant and valid for any $n$, it does not change the scaling results. 
	
	\bibliography{lib}
	\bibliographystyle{unsrt}
	
\end{document}